\documentclass[12pt,english,thmsb,titlepage]{article}
\usepackage[latin9]{inputenc}
\usepackage{amsmath}
\usepackage{amssymb}
\usepackage{stmaryrd}
\usepackage[authoryear]{natbib}
\usepackage{amsfonts}
\usepackage{babel}
\usepackage{sw20elba}

\makeatletter
\input{tcilatex}
\makeatother

\begin{document}

\author{Jose Diogo Barbosa\thanks{%
The authors gratefully acknowledge the research support of CNPq and FAPERJ.}
\\
\emph{University of Michigan}\\
\and Marcelo J. Moreira \\
\emph{FGV}}
\title{Likelihood Inference and The Role of Initial Conditions for the
Dynamic Panel Data Model}
\date{This version: \today }
\maketitle

\vspace{1in}

\begin{center}
\textbf{Abstract\medskip{} }
\end{center}

\citet{Lancaster02} proposes an estimator for the dynamic panel data model
with homoskedastic errors and zero initial conditions. In this paper, we
show this estimator is invariant to orthogonal transformations, but is
inefficient because it ignores additional information available in the data.
The zero initial condition is trivially satisfied by subtracting initial
observations from the data. We show that differencing out the data further
erodes efficiency compared to drawing inference conditional on the first
observations.

Finally, we compare the conditional method with standard random effects
approaches for unobserved data. Standard approaches implicitly rely on
normal approximations, which may not be reliable when unobserved data is
very skewed with some mass at zero values. For example, panel data on firms
naturally depend on the first period in which the firm enters on a new
state. It seems unreasonable then to assume that the process determining
unobserved data is known or stationary. We can instead make inference on
structural parameters by conditioning on the initial observations.

$\vspace{4in}$

\newpage{}

\section{Introduction}

In an important paper, \citet{Lancaster02} studies, from a Bayesian
perspective, estimation of the structural parameters of a dynamic panel data
model with fixed effects and initial observations equal to zero. His method
involves reparameterizing the model so that the information matrix is block
diagonal, with the common parameters in one block and the incidental
parameters in the other. His estimator is then defined as one of the local
maxima of the integrated likelihood function, integrating with respect to
the Lebesgue measure on $\mathbb{R}^{N}$. However, Lancaster leaves
unanswered the question of how to uniquely determine the consistent root of
his proposed methodology. Some authors, including \citet{DhaeneJochmans16}
and \citet{Kruiniger14}, have proposed different ways to find the consistent
estimator in Lancaster's approach.

In this paper, we explain the shortcoming of Lancaster's estimator: it
ignores available information in the model. In particular, Lancaster's
posterior distribution uses only part of the maximal invariant statistic's
log-likelihood function; when the full likelihood function is used in the
estimation, a unique, consistent, and asymptotically normal efficient
estimator is obtained; see \citet{Moreira09b}. The estimator obtained using
the full likelihood is asymptotically more efficient than Lancaster's
estimator. Therefore, trying to correct the nonuniqueness issue of
Lancaster's estimator is unnecessary and leads to inefficient estimators.

\citet{Lancaster02} and \citet{Moreira09b} study consistent estimation of
dynamic panel models under the same set of assumptions and with initial
observations equal to zero. The zero initial condition is trivially
satisfied when the initial observations are subtracted from the
autoregressive variables. We then show that efficiency is improved by
conditioning on the initial observations instead of differencing out the
data. The conditional argument is essentially a fixed-effects approach in
which we make no further premises on unobserved data. This is in contrast
with commonly-used estimators in the literature which make further
assumptions; see \citet{Bai13,Bai13b} on correlated random effects or %
\citet{BlundellBond98} on stationarity.

A potential advantage of conditioning on the first observation is
robustness. As \citet{BlundellSmith91} point out, asymptotic arguments are
usually based on the average temporal effect, calculated on the individual
dimension. Therefore, the importance of unobserved data does not disappear
asymptotically for relatively short panels. For example, take data on firms
or individual earnings. \citet{CabralMata03} show that the distribution of
firms is very skewed, with most of the mass being small firms, while %
\citet{Evans87, Evans87b} and \citet{Hall87} show that Gilbrat's Law
(independent firm size and growth) is rejected for small firms. As only a
handful of large firms provide the bulk of the data, we should not expect
estimators based on assumptions about unobserved data to be approximately
normal. As the firms' entry states do not disappear asymptotically for
relatively short panels, conditioning on the first observation would be
preferable to assuming known processes for unobserved data.

The remainder of this paper is organized as follows. Section 2 introduces a
simple dynamic panel data model without covariates and a zero initial
condition. This section determines the maximal invariant statistic and
summarizes the asymptotic theory for the maximum invariant likelihood
estimator (MILE). Section 3 develops the asymptotic theory for Lancaster's
estimator. Section 4 shows that this estimator is less efficient than MILE\
because it ignores relevant data. Section 5 shows that it is less efficient
to difference out the first observation than to condition on it. Section 6
compares the conditional argument to standard random effects approaches for
the unobserved data. Section 7 concludes and discusses how to extend the
model to a more useful form.

\section{The Model and the Maximal Invariant Likelihood \label{Model Sec}}

We consider a simple homoskedastic dynamic panel model with fixed effects
and without covariates:
\begin{equation}
y_{i,t+1}=\rho y_{i,t}+\eta_{i}+\sigma u_{i,t},\;i=1,\cdots,N;\;t=1,\cdots,T
\label{eq:Model}
\end{equation}
where $N\geq T+1$, $y_{i,t}\in\mathbb{R}$ are observable variables and $%
u_{i,t}\overset{iid}{\sim}N\left(0,1\right)$ are unobservable errors; $%
\eta_{i}\in\mathbb{R}$ are incidental parameters and $\left(\rho,\sigma^{2}%
\right)\in\mathbb{\mathbb{R}\times R}$ are structural parameters. We denote
the true unknown parameters by $\left(\rho^{\ast},\sigma^{\ast2},\eta_{i}^{%
\ast}\right)$. We assume that the parameter space is a compact set and $%
\sigma^{\ast2}>0$. For now, we assume that the initial \emph{observed}
condition is $y_{i,1}=0$ as \citet{Lancaster02} does. We relax this
assumption in Sections \ref{Initial Condition Sec} and \ref{Random Effects
Sec}.

Solving model (\ref{eq:Model}) recursively and writing it in matrix form
yields
\begin{eqnarray}
Y_{T} & = & \eta1_{T}^{\prime}B_{T}^{\prime}+\sigma U_{T}B_{T}^{\prime},%
\text{ where}  \label{eq:Model-Matrix-Form} \\
U_{T} & \sim & N\left(0_{N\times T},I_{N}\otimes I_{T}\right),  \notag
\end{eqnarray}
$\eta=\left(\eta_{1},\cdots,\eta_{N}\right)^{\prime}\in\mathbb{R}^{N\times1}$%
, $1_{T}=\left(1,\cdots,1\right)^{\prime}\in\mathbb{R}^{T\times1},$
\begin{equation*}
Y_{T}=\left[%
\begin{array}{ccc}
y_{1,2} & \cdots & y_{1,T+1} \\
\vdots & \ddots & \vdots \\
y_{N,2} & \cdots & y_{N,T+1}%
\end{array}%
\right],\;U_{T}=\left[%
\begin{array}{ccc}
u_{1,2} & \cdots & u_{1,T+1} \\
\vdots & \ddots & \vdots \\
u_{N,2} & \cdots & u_{N,T+1}%
\end{array}%
\right],\text{ and }B_{T}=\left[%
\begin{array}{ccc}
1 &  &  \\
\vdots & \ddots &  \\
\rho^{T-1} & \cdots & 1%
\end{array}%
\right].
\end{equation*}
When there is no confusion, we will omit the subscript from the matrices;
e.g., $Y$ instead of $Y_{T}$, $B$ instead of $B_{T}$, etc.

The inverse of $B$ has a simple form,
\begin{equation*}
B^{-1}\equiv D=I_{T}-\rho J_{T},\text{ where }J_{T}=\left[%
\begin{array}{cc}
0_{T-1}^{\prime} & 0 \\
I_{T-1} & 0_{T-1}%
\end{array}%
\right]
\end{equation*}
and $0_{T-1}$ is a $\left(T-1\right)$-dimensional column vector with zero
entries.

If individuals $i$ are treated equally, the coordinate system used to
specify the vector $\left(y_{1,t},\cdots,y_{N,t}\right)$ should not affect
inference based on them. Therefore, it is reasonable to restrict attention
to coordinate-free functions of $\left(y_{1,t},\cdots,y_{N,t}\right)$. %
\citet{ChamberlainMoreira09} and \citet{Moreira09b} show that, indeed,
orthogonal transformations preserve both the model (\ref%
{eq:Model-Matrix-Form}) and the structural parameters $\left(\rho,\sigma^{2}%
\right)$ and this yields a maximal invariant statistic, the $T\times T$
matrix $Y^{\prime}Y$. So, if the researcher finds that it is reasonable to
restrict attention to statistics that are invariant to orthogonal
transformations, the maximal invariant statistic plays a crucial role: a
statistic is invariant to orthogonal transformations if, and only if, it
depends on the data through the maximal invariant statistic $Y^{\prime}Y$.

The maximal invariant statistic $Y^{\prime}Y$ has a noncentral Wishart
distribution and depends only on $\rho$, $\sigma^{2}$, and $%
\omega_{\eta}^{2}\equiv\frac{\eta^{\prime}\eta}{\sigma^{2}N}$. The
noncentral Wishart distribution is the multivariate generalization of the
noncentral Chi-squared distribution and it depends on the modified Bessel
function of the first kind. We use uniform approximations of Bessel
functions (see \citet{AbramowitzStegun65}), which allows us to write the
density of the noncentral Wishart distribution in a more tractable form.
Specifically, the log-likelihood of $Y^{\prime}Y$ is, up to an $%
o_{p}\left(N^{-1}\right)$ term, proportional to
\begin{eqnarray}
Q_{N}^{M}\left(\theta\right) & = & -\frac{1}{2}\ln\left(\sigma^{2}\right)-%
\frac{1}{2\sigma^{2}}\frac{tr\left(DY^{\prime}YD^{\prime}\right)}{NT}-\frac{%
\omega_{\eta}^{2}}{2}  \notag \\
& & +\frac{\left(1+A_{N}^{2}\right)^{1/2}}{2T}-\frac{\ln\left(1+%
\left(1+A_{N}^{2}\right)^{1/2}\right)}{2T},
\label{eq:Objetive-Function-MILE1}
\end{eqnarray}
where $\theta=\left(\rho,\sigma^{2},\omega_{\eta}^{2}\right)$ and $A_{N}=2%
\sqrt{\omega_{\eta}^{2}\frac{1_{T}^{\prime}DY^{\prime}YD^{\prime}1_{T}}{%
\sigma^{2}N}}$.

The log-likelihood $Q_{N}^{M}\left( \theta \right) $ is free from the
incidental parameter problem since $Y^{\prime }Y$ is parametrized by the
fixed dimensional vector of parameters $\theta $. Although the dimension is
fixed, the parameter
\begin{equation}
\omega _{\eta }^{2}\equiv \frac{\eta ^{\prime }\eta }{\sigma ^{2}N}=\frac{%
\sum_{i=1}^{N}\eta _{i}^{2}}{\sigma ^{2}N}
\end{equation}%
depends on the sample size $N$. For simplicity, we omit the dependence on $N$
from the parameter $\omega _{\eta }^{2}$. However, the asymptotic properties
of the estimator will be derived under different sequences of $\omega _{\eta
}^{2}$. The asymptotic properties of the estimator obtained by maximizing
the objective function $Q_{N}^{M}\left( \theta \right) $ are studied by %
\citet{Moreira09b} and we reproduce these results here for convenience. The
information matrix $\mathcal{I}_{T}\left( \theta ^{\ast }\right) $ in %
\citet{Moreira09b} contains typographical errors which are corrected here.
Define the matrix
\begin{equation*}
F_{0}=\left[
\begin{array}{ccccc}
0 & 0 & \cdots & 0 & 0 \\
\rho & 0 & \cdots & 0 & 0 \\
\frac{1}{2}\rho ^{2} & \rho & \cdots & 0 & 0 \\
\vdots & \vdots & \ddots & \vdots & \vdots \\
\frac{1}{T-1}\rho ^{T-1} & \frac{1}{T-2}\rho ^{T-2} & \cdots & \rho & 0%
\end{array}%
\right]
\end{equation*}%
and its derivatives:
\begin{equation*}
F_{j}=\frac{d^{j}}{d\rho ^{j}}F_{0}\text{ and }F_{j}^{\ast }=\left. \frac{%
d^{j}}{d\rho ^{j}}F_{0}\right\vert _{\rho =\rho ^{\ast }}.
\end{equation*}

\bigskip{}

{\large {}{}{}{}{}{}Theorem 1}: Let
\begin{equation}
\hat{\theta}=\underset{\theta\in\varTheta}{\arg\max}Q_{N}^{M}\left(\theta%
\right).  \label{eq:MILE}
\end{equation}

(A.1) Under the assumption that $N\rightarrow\infty$ with $T$ fixed, (i) if $%
\omega_{\eta}^{2}$ is fixed at $\omega_{\eta}^{\ast2}$, then $\hat{\theta}%
_{M}\rightarrow_{p}\theta^{\ast}=\left(\rho^{\ast},\sigma^{\ast2},\omega_{%
\eta}^{2\ast}\right)$; (ii) if $\omega_{\eta}^{2}\rightarrow\omega_{\eta}^{%
\ast2}$, then $\hat{\theta}_{M}\rightarrow_{p}\theta^{\ast}=\left(\rho^{%
\ast},\sigma^{\ast2},\omega_{\eta}^{\ast2}\right)$; and (iii) if $%
\limsup\omega_{\eta}^{\ast2}<\infty$, then $\hat{\theta}_{M}=\theta^{%
\ast}+o_{p}(1)$, where $\theta^{\ast}=\left(\rho^{\ast},\sigma^{\ast2},%
\omega_{\eta}^{\ast2}\right)$.

(A.2) Under the assumption that $T\rightarrow\infty$ and $\left\vert
\rho^{\ast}\right\vert <1$, (i) if $\omega_{\eta}^{2}$ is fixed at $%
\omega_{\eta}^{\ast2}$, then $\hat{\theta}_{M}\rightarrow_{p}\theta^{\ast}=%
\left(\rho^{\ast},\sigma^{\ast2},\omega_{\eta}^{\ast2}\right)$; (ii) if $%
\omega_{\eta}^{2}\rightarrow\omega_{\eta}^{\ast2}$, then $\hat{\theta}%
_{M}\rightarrow_{p}\theta^{\ast}=\left(\rho^{\ast},\sigma^{\ast2},\omega_{%
\eta}^{\ast2}\right)$; and (iii) if $\limsup\omega_{\eta}^{\ast2}<\infty$,
then $\hat{\theta}_{M}=\theta^{\ast}+o_{p}(1)$, where $\theta^{\ast}=\left(%
\rho^{\ast},\sigma^{\ast2},\omega_{\eta}^{\ast2}\right)$.

(B) Assume that $\omega_{\eta}^{\ast2}>0$ is fixed, and let the score
statistic and the Hessian matrix be
\begin{equation*}
S_{N}^{M}\left(\theta\right)=\frac{\partial Q_{N}^{M}\left(\theta\right)}{%
\partial\theta}\;\text{and }H_{N}^{M}\left(\theta\right)=\frac{%
\partial^{2}Q_{N}^{M}\left(\theta\right)}{\partial\theta\partial\theta^{%
\prime}},
\end{equation*}
respectively, and define the matrix
\begin{equation*}
\mathcal{I}_{T}^{M}\left(\theta^{\ast}\right)=\left[%
\begin{array}{ccc}
h_{T}^{M} & \frac{\omega_{\eta}^{\ast4}}{\sigma^{\ast2}}\frac{%
1_{T}^{\prime}F_{1}^{\ast}1_{T}}{1+2\omega_{\eta}^{\ast2}T} & \frac{%
1+\omega_{\eta}^{\ast2}T}{1+2\omega_{\eta}^{\ast2}T}\frac{%
1_{T}^{\prime}F_{1}^{\ast}1_{T}}{T} \\
\frac{\omega_{\eta}^{\ast4}}{\sigma^{\ast2}}\frac{1_{T}^{\prime}F_{1}^{%
\ast}1_{T}}{1+2\omega_{\eta}^{\ast2}T} & \frac{1}{2\left(\sigma^{\ast2}%
\right)^{2}}\left(1+\frac{\omega_{\eta}^{\ast4}T}{1+2\omega_{\eta}^{\ast2}T}%
\right) & \frac{1}{2\sigma^{\ast2}}\frac{1+\omega_{\eta}^{\ast2}T}{%
1+2\omega_{\eta}^{\ast2}T} \\
\frac{1+\omega_{\eta}^{\ast2}T}{1+2\omega_{\eta}^{\ast2}T}\frac{%
1_{T}^{\prime}F_{1}^{\ast}1_{T}}{T} & \frac{1}{2\sigma^{\ast2}}\frac{%
1+\omega_{\eta}^{\ast2}T}{1+2\omega_{\eta}^{\ast2}T} & \frac{T}{%
2\left(1+2\omega_{\eta}^{\ast2}T\right)}%
\end{array}%
\right],
\end{equation*}
where
\begin{equation*}
h_{T}^{M}=\frac{tr\left(F_{1}^{\ast}F_{1}^{^{\prime}\ast}\right)}{T}+\frac{%
\omega_{\eta}^{\ast4}T}{1+\omega_{\eta}^{\ast2}T}\left(\frac{%
1_{T}^{\prime}F_{1}^{\ast}F_{1}^{^{\prime}\ast}1_{T}}{T}+\frac{1}{%
1+2\omega_{\eta}^{\ast2}T}\left(\frac{1_{T}^{\prime}F_{1}^{\ast}1_{T}}{T}%
\right)^{2}\right).
\end{equation*}
As $N\rightarrow\infty$ with $T$ fixed, (i) $\sqrt{NT}S_{N}^{M}\left(%
\theta^{\ast}\right)\rightarrow_{d}N\left(0,\mathcal{I}_{T}^{M}\left(%
\theta^{\ast}\right)\right)$; (ii) $H_{N}^{M}\left(\theta^{\ast}\right)%
\rightarrow_{p}-\mathcal{I}_{T}^{M}\left(\theta^{\ast}\right)$; (iii) $\sqrt{%
NT}\left(\hat{\theta}_{M}-\theta^{\ast}\right)\rightarrow_{d}N\left(0,%
\mathcal{I}_{T}^{M}\left(\theta^{\ast}\right)^{-1}\right)$; and (iv) the
log-likelihood ratio is
\begin{eqnarray*}
\Lambda_{N}\left(\theta^{\ast}+h\cdot\left(NT\right)^{-1/2},\theta^{\ast}%
\right) & = &
NT\left(Q_{N}^{M}\left(\theta^{\ast}+h\cdot\left(NT\right)^{-1/2}%
\right)-Q_{N}^{M}\left(\theta^{\ast}\right)\right) \\
& = & h^{\prime}\sqrt{NT}S_{N}^{M}\left(\theta^{\ast}\right)-\frac{1}{2}%
h^{\prime}\mathcal{I}_{T}\left(\theta^{\ast}\right)h+o_{Q_{N}^{M}\left(%
\theta^{\ast}\right)}(1),
\end{eqnarray*}
$\sqrt{NT}S_{N}^{M}\left(\theta^{\ast}\right)\rightarrow_{d}N\left(0,%
\mathcal{I}_{T}^{M}\left(\theta^{\ast}\right)\right)$ under $%
Q_{N}^{M}\left(\theta^{\ast}\right)$. Furthermore, $\hat{\theta}_{M}$ is
asymptotically efficient within the class of regular invariant estimators
for the differenced model (\ref{eq:Dif-Model}) under large $N$, fixed $T$
asymptotics.

\bigskip{}

Part (A) of the above theorem implies that $\hat{\rho}_{M}\rightarrow_{p}%
\rho^{\ast}$ and $\hat{\sigma}_{M}^{2}\rightarrow_{p}\sigma^{\ast2}$
regardless of the growth rate of $N$ and $T$ as long as $NT\rightarrow\infty$%
. Part (B) derives the limiting distribution of $\hat{\theta}_{M}$. It
shows, in particular, that $\hat{\rho}_{M}$ achieves the efficiency bound $%
\left(\mathcal{I}_{T}^{M}\left(\theta^{\ast}\right)^{-1}\right)_{11}$ for
regular invariant estimators as $N\rightarrow\infty$. Regular estimators
exclude superefficient estimators in the sense of Hodges-Le Cam and,
heuristically, a regular estimator is one whose asymptotic distribution does
not change in shrinking neighborhoods of the true parameter value (see %
\citet{BickelKlaasenRitovWellner} for more details).

Part (B) of Theorem 1 finds the asymptotic distribution of $\hat{\theta}_{M}$%
, assuming that $\omega_{\eta}^{2}$ is fixed at $\omega_{\eta}^{\ast2}$. A
generalization of part (B) that allows for $\omega_{\eta}^{2}\rightarrow%
\omega_{\eta}^{\ast2}$ follows from a simple application of Le Cam's third
lemma.

\bigskip{}

{\large {}{}{}{}{}{}Lemma 2}: Assume $\omega_{\eta}^{2}=\omega_{\eta}^{%
\ast2}+h/\sqrt{N}$, where $\omega_{\eta}^{\ast2}>0$, $h\in\mathbb{R}$ and $%
\omega_{\eta}^{\ast2}$ is in the parameter space. As $N\rightarrow\infty$
with $T$ fixed,
\begin{equation*}
\sqrt{NT}\left(\hat{\theta}_{M}-\theta^{\ast}\right)\rightarrow_{d}N\left(%
\left(%
\begin{array}{c}
0 \\
0 \\
h%
\end{array}%
\right),\mathcal{I}_{T}^{M}\left(\theta^{\ast}\right)^{-1}\right),
\end{equation*}
where $\mathcal{I}_{T}^{M}\left(\theta^{\ast}\right)^{-1}$ is defined in
Theorem 1, part (B).

\bigskip{}

Lemma 2 shows that the asymptotic distribution of the structural parameters
does not change, whether $\omega_{\eta}^{2}$ is fixed at $%
\omega_{\eta}^{\ast2}$ or $\omega_{\eta}^{2}\rightarrow\omega_{\eta}^{\ast2}$%
. For simplicity's sake, we assume throughout the paper that the sequence $%
\omega_{\eta}^{2}$ (which can depend on the sample size $N$) is fixed at $%
\omega_{\eta}^{\ast2}>0$.

\section{Lancaster's Estimator \label{Lancaster Estimator Sec}}

\citet{Lancaster02} proposes a Bayesian approach to estimate the structural
parameters which involves reparameterizing model \eqref{eq:Model-Matrix-Form}
so that the information matrix is block diagonal, with the common parameters
in one block and the incidental parameters in the other. Then he defines his
estimator as one of the local maxima of the integrated likelihood function,
integrating with respect to the Lebesgue measure on $\mathbb{R}^{N}$. %
\citet{DhaeneJochmans16} give an alternative interpretation for Lancaster's
estimator by showing that Lancaster's posterior can be obtained by adjusting
the profile likelihood so that its score is free from asymptotic bias. In
this sense, Lancaster's estimator is a bias-corrected estimator.

Lancaster's estimator seeks to maximize the following objective function:
\begin{equation}
Q_{N}^{L}\left(\rho,\sigma^{2}\right)=-\frac{1}{2}\ln\left(\sigma^{2}\right)+%
\frac{1_{T}^{\prime}F_{0}1_{T}}{T\left(T-1\right)}-\frac{1}{2\sigma^{2}}%
\frac{tr\left(DY^{\prime}YD^{\prime}H\right)}{N\left(T-1\right)},
\label{eq:Lancaster-Object-1}
\end{equation}
where the matrix $H$ is defined as
\begin{equation*}
H=I_{T}-\frac{1}{T}1_{T}1_{T}^{\prime}.
\end{equation*}

The posterior (\ref{eq:Lancaster-Object-1}) is not a likelihood function and
$\left(\rho^{\ast},\sigma^{\ast2}\right)$ is not a global maximizer of $%
\underset{N\rightarrow\infty}{plim}Q_{N}^{L}\left(\rho,\sigma^{2}\right)$ as
in standard maximum likelihood theory (see \citet{DhaeneJochmans16}).
Nonetheless, Theorem 3, proved by \citet{Lancaster02}, shows that the
posterior (\ref{eq:Lancaster-Object-1}) can be used to consistently estimate
the structural parameters.

\bigskip{}

{\large {}{}{}{}{}{}Theorem 3}: Let $S_{N}^{L}\left(\rho,\sigma^{2}\right)=%
\frac{\partial Q_{N}^{L}\left(\rho,\sigma^{2}\right)}{\partial\left(\rho,%
\sigma^{2}\right)^{\prime}}$ be the score of the posterior (\ref%
{eq:Lancaster-Object-1}) and let $\Theta_{N}$ be the set of roots of
\begin{equation}
S_{N}^{L}\left(\rho,\sigma^{2}\right)=0  \label{eq:Lancaster-Estimation-Eq}
\end{equation}
corresponding to the local maxima. If that set is empty, set $\Theta_{N}$
equals to $\left\{ 0\right\} $. Then, there is a consistent root of equation
(\ref{eq:Lancaster-Estimation-Eq}).

\bigskip{}

The usefulness of Theorem 3 is limited by the fact that it only states that
one (of possibly many) of the local maxima of (\ref{eq:Lancaster-Object-1})
is a consistent estimator of the structural parameters. However, it does not
indicate how to find a consistent estimator. Therefore, even though the
posterior can be used to consistently estimate the structural parameters of
model \eqref{eq:Model-Matrix-Form}, Lancaster leaves unanswered the question
of how to uniquely determine the consistent root of (\ref%
{eq:Lancaster-Estimation-Eq}). To our knowledge, two different methodologies
to uniquely choose the consistent root of Lancaster's score have been
proposed. The first approach, by \citet{DhaeneJochmans16}, suggests as a
consistent estimator of the structural parameters, the minimizer of the norm
of Lancaster's score on an interval around the maximum likelihood estimator
obtained from the maximization of the likelihood of %
\eqref{eq:Model-Matrix-Form}. The second method, by \citet{Kruiniger14},
uses as a consistent estimator, the minimizer of a quadratic form in
Lancaster's score subjected to a condition on the Hessian matrix.

Lemma 4 finds the asymptotic variance of a consistent root of (\ref%
{eq:Lancaster-Estimation-Eq}).

\bigskip{}

{\large {}{}{}{}{}{}Lemma 4}: Assume $\omega_{\eta}^{\ast2}>0$ and let $%
\left(\hat{\rho}_{L},\hat{\sigma}_{L}^{2}\right)$ be a consistent root of $%
S_{N}^{L}\left(\rho,\sigma^{2}\right)$. Let the Hessian matrix be

\begin{equation*}
H_{N}^{L}\left(\rho,\sigma^{2}\right)=\frac{\partial^{2}Q_{N}^{L}\left(\rho,%
\sigma^{2}\right)}{\partial\left(\rho,\sigma^{2}\right)^{\prime}\partial%
\left(\rho,\sigma^{2}\right)},
\end{equation*}
and define the matrices
\begin{equation*}
\mathcal{I}_{T}^{L}\left(\theta^{*}\right)=\left[%
\begin{array}{cc}
h_{T}^{L} & -\frac{1}{\sigma^{*2}}\frac{1_{T}^{\prime}F_{1}^{*}1_{T}}{%
T\left(T-1\right)} \\
-\frac{1}{\sigma^{*2}}\frac{1_{T}^{\prime}F_{1}^{*}1_{T}}{T\left(T-1\right)}
& \frac{1}{2\left(\sigma^{*2}\right)^{2}}%
\end{array}%
\right]
\end{equation*}
and
\begin{equation}
\Sigma_{T}\left(\theta^{*}\right)=\frac{T}{T-1}\left[%
\begin{array}{cc}
a_{T}^{L} & \mathcal{I}_{T}^{L}\left(1,2\right) \\
\mathcal{I}_{T}^{L}\left(1,2\right) & \mathcal{I}_{T}^{L}\left(2,2\right)%
\end{array}%
\right],  \label{eq:Var-Score-lancaster}
\end{equation}

where
\begin{equation*}
h_{T}^{L}=-\frac{1_{T}^{\prime}F_{2}^{\ast}1_{T}}{T\left(T-1\right)}+\frac{%
tr\left(F_{1}^{\ast}F_{1}^{\ast^{\prime}}\right)}{T-1}+\frac{%
1_{T}^{\prime}F_{1}^{\ast^{\prime}}F_{1}^{\ast}1_{T}}{T}\frac{%
\left(\omega_{\eta}^{\ast2}T-1\right)}{T-1}-\frac{\omega_{\eta}^{\ast2}T}{T-1%
}\left(\frac{1_{T}^{\prime}F_{1}^{\ast^{\prime}}1_{T}}{T}\right)^{2},
\end{equation*}
\begin{equation*}
a_{T}^{L}=2h_{T}^{L}+2\frac{1_{T}^{\prime}F_{2}^{\ast}1_{T}}{%
T\left(T-1\right)},
\end{equation*}
and $\mathcal{I}_{T}^{L}\left(i,j\right)$ is the $\left(i,j\right)$-th entry
of $\mathcal{I}_{T}^{L}\left(\theta^{\ast}\right)$. When $\mathcal{I}%
_{T}^{L}\left(\theta^{\ast}\right)$ is nonsingular, as $N\rightarrow\infty$
with $T$ fixed, (i) $\sqrt{NT}S_{N}^{L}\left(\rho^{\ast},\sigma^{\ast2}%
\right)\rightarrow_{d}N\left(0,\Sigma_{T}\left(\theta^{\ast}\right)\right)$;
(ii) $-H_{N}^{L}\left(\rho^{\ast},\sigma^{\ast2}\right)\rightarrow_{p}%
\mathcal{I}_{T}^{L}\left(\theta^{\ast}\right)$; and (iii) $\sqrt{NT}\left(%
\begin{array}{c}
\hat{\rho}_{L}-\rho^{\ast} \\
\hat{\sigma^{2}}_{L}-\sigma^{\ast2}%
\end{array}%
\right)\rightarrow_{d}N\left(0,\mathcal{I}_{T}^{L}\left(\theta^{\ast}%
\right)^{-1}\Sigma_{T}\left(\theta^{\ast}\right)\mathcal{I}%
_{T}^{L}\left(\theta^{\ast}\right)^{-1}\right)$.

\bigskip{}

Lemma 4 follows from standard asymptotic theory because of the assumption
that the Hessian matrix $-H_{N}^{L}\left(\rho^{\ast},\sigma^{\ast2}\right)$
converges in probability to a nonsingular matrix $\mathcal{I}%
_{T}^{L}\left(\theta^{\ast}\right)$. \citet{DhaeneJochmans16} prove that
this assumption is violated, i.e. $\mathcal{I}_{T}^{L}\left(\theta^{\ast}%
\right)$ is singular, when $T=2$ or $\rho^{\ast}=1$.

\section{Lancaster's Estimator and a Decomposition of the Maximal Invariant
Statistic's Log-Likelihood \label{Decomposition Sec}}

In contrast to Lancaster's estimator, the estimator defined in (\ref{eq:MILE}%
), based on the maximal invariant statistic's log-likelihood, is a standard
maximum likelihood estimator in the sense that its objective function is a
likelihood function and the limit of its objective function attains a unique
global maximum at the vector of true parameters. This implies that the
estimator that maximizes the maximal invariant statistic's log-likelihood is
uniquely determined and no additional procedures are necessary to obtain a
consistent and asymptotically normal estimator of the structural parameters.
More than that, Theorem 1 shows that it attains the minimum variance bound
for invariant regular estimators of the structural parameters of model %
\eqref{eq:Model-Matrix-Form}. Therefore, the estimator defined in (\ref%
{eq:MILE}) efficiently uses all information available in the maximal
invariant statistic $Y^{\prime}Y$.

Lancaster's estimator is also invariant to orthogonal transformations since
it depends on the data only through the maximal invariant statistic $%
Y^{\prime}Y$; see equation (\ref{eq:Lancaster-Object-1}). However, it is not
uniquely determined and additional steps, such as those proposed by %
\citet{DhaeneJochmans16} and \citet{Kruiniger14}, are necessary to choose
the consistent root among the (possibly) many roots of Lancaster's score
function. Furthermore, the limit of probability of the Hessian of $%
Q_{N}^{L}\left(\rho,\sigma^{2}\right)$ is not necessarily proportional to
the asymptotic variance (\emph{AVar}) of the score. Since the information
equality does not hold, the asymptotic variance of Lancaster's consistent
estimator cannot attain the lower bound for invariant regular estimators
found in Theorem 1.

The explanation for the shortcomings of Lancaster's estimator is that, even
though it is invariant, its objective function ignores information available
in the maximal invariant by using only part of the maximal invariant
statistic's log-likelihood function. Indeed, the maximal invariant
statistic's log-likelihood function (\ref{eq:Objetive-Function-MILE1})
concentrated out of $\sigma^{2}$ and $\omega_{\eta}^{2}$ is
\begin{equation}
Q_{N}^{M}\left(\rho\right)=-\frac{1}{2}\frac{T-1}{T}\ln\left(\frac{%
tr\left(DY^{\prime}YD^{\prime}H\right)}{N\left(T-1\right)}\right)-\frac{1}{2T%
}\ln\left(\frac{1_{T}^{\prime}DY^{\prime}YD^{\prime}1_{T}}{NT}\right),
\label{eq:MILE-Concentrated1}
\end{equation}
while Lancaster's objective function concentrated out of $\sigma^{2}$ is
\begin{equation*}
Q_{N}^{L}\left(\rho\right)=\frac{1_{T}^{\prime}F_{0}1_{T}}{T\left(T-1\right)}%
-\frac{1}{2}\ln\left(\frac{tr\left(DY^{\prime}YD^{\prime}H\right)}{%
N\left(T-1\right)}\right),
\end{equation*}
which allows us to write
\begin{eqnarray}
Q_{N}^{M}\left(\rho\right) & = & \frac{T-1}{T}Q_{N}^{L}\left(\rho%
\right)+Q_{N}^{M-L}\left(\rho\right),\text{ where}  \label{eq:Decomposition1}
\\
Q_{N}^{M-L}\left(\rho\right) & = & -\frac{1_{T}^{\prime}F_{0}1_{T}}{T^{2}}-%
\frac{1}{2T}\ln\left(\frac{1_{T}^{\prime}DY^{\prime}YD^{\prime}1_{T}}{NT}%
\right).  \notag
\end{eqnarray}
The respective score functions are given by
\begin{equation*}
S_{N}^{L}\left(\rho\right)=\frac{1_{T}^{\prime}F_{1}1_{T}}{T\left(T-1\right)}%
+\frac{tr\left(J_{T}Y^{\prime}YD^{\prime}H\right)}{tr\left(DY^{\prime}YD^{%
\prime}H\right)}
\end{equation*}
and
\begin{eqnarray}
S_{N}^{M}\left(\rho\right) & = & \frac{T-1}{T}S_{N}^{L}\left(\rho%
\right)+S_{N}^{M-L}\left(\rho\right),\text{ where} \\
S_{N}^{M-L}\left(\rho\right) & = & -\frac{1_{T}^{\prime}F_{1}1_{T}}{T^{2}}+%
\frac{1}{T}\frac{1_{T}^{\prime}J_{T}Y^{\prime}YD^{\prime}1_{T}}{%
1_{T}^{\prime}DY^{\prime}YD^{\prime}1_{T}}.  \notag
\end{eqnarray}
The decomposition (\ref{eq:Decomposition1}) implies an orthogonal
decomposition of the concentrated score function:

\bigskip{}

{\large {}{}{}{}{}{}Theorem 5}: Let $S_{N}^{M}\left(\rho\right)$, $%
S_{N}^{L}\left(\rho\right)$ and $S_{N}^{M-L}\left(\rho\right)$ be the score
functions associated with $Q_{N}^{M}\left(\rho\right)$, $Q_{N}^{L}\left(\rho%
\right)$ and $Q_{N}^{M-L}\left(\rho\right)$, respectively. Then,
\begin{equation}
S_{N}^{M}\left(\rho\right)=\frac{T-1}{T}S_{N}^{L}\left(\rho%
\right)+S_{N}^{M-L}\left(\rho\right),  \label{eq:Decomposition2}
\end{equation}
where
\begin{equation}
S_{N}^{M-L}\left(\rho^{\ast}\right)\rightarrow_{p}0
\label{eq:Limit-Score-M-L}
\end{equation}
and
\begin{equation}
\left(%
\begin{array}{c}
\sqrt{NT}S_{N}^{L}\left(\rho^{\ast}\right) \\
\sqrt{NT}S_{N}^{M-L}\left(\rho^{\ast}\right)%
\end{array}%
\right)\rightarrow_{d}N\left(\left(%
\begin{array}{c}
0 \\
0%
\end{array}%
\right),\left(%
\begin{array}{cc}
b_{T} & 0 \\
0 & c_{T}%
\end{array}%
\right)\right),  \label{eq:Scores-concentrated-Distribution}
\end{equation}
with
\begin{equation*}
b_{T}=a_{T}^{L}-2\frac{T}{\left(T-1\right)^{3}}\left(\frac{%
1_{T}^{\prime}F_{1}^{\ast}1_{T}}{T}\right)^{2}>0,
\end{equation*}
the constant $a_{T}^{L}$ as defined in Lemma 4, and
\begin{equation*}
c_{T}=\frac{2T}{\left(1+\omega_{\eta}^{\ast2}T\right)}\left(\frac{%
1_{T}^{\prime}F_{1}^{\ast^{\prime}}F_{1}^{\ast}1_{T}}{T}-\left(\frac{%
1_{T}^{\prime}F_{1}^{\ast}1_{T}}{T}\right)^{2}\right)>0.
\end{equation*}

\bigskip{}

A consistent estimator for $\rho ^{\ast }$ is expected to be more efficient,
the closer it stays to the maximum likelihood estimator $\hat{\rho}_{M}$.
This implies that the consistent estimator $\hat{\rho}_{L}$ might have low
efficiency because it only uses part of the full log-likelihood (\ref%
{eq:Decomposition1}). One may use $\left\vert S_{N}^{M}\left( \hat{\rho}%
_{L}\right) \right\vert $ as a measure of how \textquotedblleft
close\textquotedblright\ any inflection point $\hat{\rho}_{L}$ is to $\hat{%
\rho}_{M}$ and, because $S_{N}^{M}\left( \hat{\rho}_{L}\right)
=S_{N}^{M-L}\left( \hat{\rho}_{L}\right) $, $\hat{\rho}_{L}$ will be more
efficient, the smaller $\left\vert S_{N}^{M-L}\left( \hat{\rho}_{L}\right)
\right\vert $ is. The term $S_{N}^{M-L}\left( .\right) $ evaluated at the
true value $\rho ^{\ast }$ or at a consistent estimator, converges in
probability to zero. This suggests choosing the inflexion points based on
Lancaster's score by looking at $S_{N}^{M-L}\left( \hat{\rho}_{L}\right) $
to uniquely determine the consistent root, akin to \citet{DhaeneJochmans16}
and \citet{Kruiniger14}.

Instead, we can use the information on $S_{N}^{M-L}\left(.\right)$ to
increase the efficiency of $\hat{\rho}_{L}$. Simple algebra shows that
\begin{equation*}
\left\vert S_{N}^{M-L}\left(\hat{\rho}_{L}\right)\right\vert =\left\vert
\left(T-1\right)\frac{tr\left(J_{T}Y^{\prime}Y\hat{D}_{L}^{\prime}H\right)}{%
tr\left(\hat{D}_{L}Y^{\prime}Y\hat{D}_{L}^{\prime}H\right)}+\frac{%
1_{T}^{\prime}J_{T}Y^{\prime}Y\hat{D}_{L}^{\prime}1_{T}}{1_{T}^{\prime}\hat{D%
}_{L}Y^{\prime}Y\hat{D}_{L}^{\prime}1_{T}}\right\vert ,
\end{equation*}
where $\hat{D}_{L}$ is the matrix $D$ evaluated at $\hat{\rho}_{L}$. Notice
that $\hat{\rho}_{L}$ is inefficient since $\sqrt{NT}S_{N}^{M-L}\left(\hat{%
\rho}_{L}\right)$ does not converge in probability to zero. Instead of
choosing the consistent root following Lancaster's methodology, we could add
the information in $S_{N}^{M-L}\left(.\right)$ to $S_{N}^{L}\left(.\right)$
by using (\ref{eq:Decomposition2}). The associated estimator $\hat{\rho}_{M}$
uses the information in both $S_{N}^{M-L}\left(.\right)$ and $%
S_{N}^{L}\left(.\right)$ efficiently and is uniquely defined, making the
estimation problem more simple and objective.

\section{Initial Condition and Likelihood Inference\label{Initial Condition
Sec}}

Theorem 1 shows that the estimator for the structural parameters $\left(
\rho ,\sigma ^{2}\right) $ attains the efficiency bound for model (\ref%
{eq:Model}) when the first observation is $y_{i,1}=0$. \citet{Lancaster02}
suggests working with the differenced data $\tilde{y}_{i,t}=y_{i,t}-y_{i,1}$
so that we may continue working with an initial observation equal to zero.
Instead, we draw inference on $\left( \rho ,\sigma ^{2}\right) $ by
conditioning on the first observation $y_{i,1}$. We will show the
differencing method is indeed less efficient than the conditional method.

By using model (\ref{eq:Model}) with the variables $y_{i,t}$ in levels,
there are more observations that can be used to estimate the parameters. On
the other hand, the number of parameters to be estimated increases. In this
section we show that the estimator that maximizes the likelihood of the
maximal invariant statistic of the original model (\ref{eq:Model}) has
asymptotic variance that is no larger than the variance of the estimator (%
\ref{eq:MILE}) and, under very general conditions, using $y_{i,t}$ in levels
will be strictly more efficient.

When we allow for a nonzero initial condition, the model \eqref{eq:Model}
becomes
\begin{eqnarray}
Y_{T} & = & y_{1}.\rho
e_{1}^{\prime}B_{T}^{\prime}+\eta.1_{T}^{\prime}B^{\prime}+\sigma.U_{T}.B^{%
\prime},\text{ where}  \label{eq:Initial-Model} \\
U_{T} & \sim & N\left(0_{N\times T},I_{N}\otimes I_{T}\right),  \notag
\end{eqnarray}
$y_{1}=$ $\left(y_{1,1},\cdots,y_{N,1}\right)^{\prime}$ is the vector of
first observations and $e_{1}=\left(1,0,...,0\right)^{\prime}$ is the
canonical vector.

\citet{Lancaster02} works with a differenced version of model (\ref{eq:Model}%
) given by
\begin{eqnarray}
\widetilde{y}_{i,t+1} & = & \rho\widetilde{y}_{i,t}+\widetilde{\eta}%
_{i}+\sigma u_{i,t},\;i=1,\cdots,N;\;t=1,\cdots,T  \label{eq:Dif-Model} \\
\widetilde{y}_{i,t} & \equiv &
y_{i,t}-y_{i,1},\;i=1,\cdots,N;\;t=1,\cdots,T+1  \notag \\
\widetilde{\eta}_{i} & \equiv &
\eta_{i}-y_{i,1}\left(1-\rho\right),\;i=1,\cdots,N  \notag
\end{eqnarray}
and seeks inference based on $\widetilde{y}_{i,1}=0$ for all $i$. In matrix
form,

\begin{equation*}
Y_{T+1}-y_{1}.1_{T+1}^{\prime}=\left[0_{N\times1}:Y_{T}-y_{1}.1_{T}^{\prime}%
\right].
\end{equation*}
The second term equals
\begin{eqnarray*}
Y_{T}-y_{1}.1_{T}^{\prime} & = &
y_{1}.\left(\rho.e_{1}^{\prime}B^{\prime}-1_{T}^{\prime}\right)+\eta.1_{T}^{%
\prime}B^{\prime}+\sigma.U.B^{\prime} \\
& = & \left[\eta-\left(1-\rho\right)y_{1}\right].1_{T}^{\prime}B^{\prime}+%
\sigma.U.B^{\prime}.
\end{eqnarray*}

Defining differenced variables and the incidental parameters,
\begin{equation*}
\widetilde{Y}_{T}=Y_{T}-y_{1}.1_{T}^{\prime}\text{ and }\widetilde{\eta}%
=\eta-\left(1-\rho\right).y_{1},
\end{equation*}
we are back to the model with the first observation equal to zero:
\begin{equation}
\widetilde{Y}_{T}=\widetilde{\eta}.1_{T}^{\prime}B^{\prime}+\sigma.U_{T}.B^{%
\prime},\;U\sim N\left(0_{N\times T},I_{N}\otimes I_{T}\right),
\label{eq:Model-1-Matrix-Form}
\end{equation}
where the incidental parameter is $\tau=\widetilde{\eta}$.

Differencing eliminates one time period from the data. Instead, we will
condition on the initial observation $y_{1}$ itself. It is convenient to
work with the linear setup of \citet{ChamberlainMoreira09}:
\begin{equation}
Y=x.a\left( \gamma \right) +\tau .b\left( \gamma \right) +U.c\left( \gamma
\right) ,  \label{eq:General-Model}
\end{equation}%
where $x\in \mathbb{R}^{N\times K}$ are explanatory variables, $\tau \in
\mathbb{R}^{N\times J}$ are the incidental parameters, and $a\left( \cdot
\right) ,$ $b\left( \cdot \right) ,$ $c\left( \cdot \right) $ are given
functions of the unknown parameter of interest $\gamma $. Consider the group
of transformations $g.Y$, where $g$ are orthogonal matrices such that $g.x=x$%
. This group modifies the unknown incidental parameter $\tau $, but
preserves the model and the parameter $\gamma $ of interest:
\begin{eqnarray*}
g.Y &=&g.x.a\left( \gamma \right) +g.\tau .b\left( \gamma \right)
+g.U.c\left( \gamma \right) \\
&=&x.a\left( \gamma \right) +\left( g.\tau \right) .b\left( \gamma \right)
+U.c\left( \gamma \right)
\end{eqnarray*}%
in distribution, because the law of $g.U$ is the same as the law of $U$.
This group yields the maximal invariant statistic, which is given by the
pair
\begin{equation}
Z_{1}=\left( x^{\prime }x\right) ^{-1/2}x^{\prime }Y\text{ and }%
Z_{2}^{\prime }Z_{2}=Y^{\prime }M_{x}Y,  \label{eq:Maximal-Invariant-1}
\end{equation}%
where $Z_{2}=q_{2}^{\prime }Y$ for any matrix $q_{2}$ such that the matrix $%
q=\left[ x\left( x^{\prime }x\right) ^{-1/2}:q_{2}\right] $ is orthogonal.
Because $q$ is an orthogonal matrix, we have $q_{2}q_{2}^{\prime
}=I_{N}-N_{x}=M_{x}$ for $N_{x}=x\left( x^{\prime }x\right) ^{-1}x^{\prime }$%
. We refer the reader to \citet{ChamberlainMoreira09} for more details.

Define the parameters
\begin{equation}
\omega_{\tau,x}^{2}=\frac{\tau^{\prime}M_{x}\tau}{\sigma^{2}N}\text{ and }%
\delta_{\tau,x}=\left(x^{\prime}x\right)^{-1}x^{\prime}\tau.
\end{equation}
The coefficient $\delta_{\tau,x}$ is the ordinary least squares (OLS)\
estimator and $\omega_{\tau,x}^{2}$ is the standardized average of sum of
squared residuals from a fictitious regression of $\tau$ on $x$ (if there
are no covariates $x$, we define $\omega_{\tau}^{2}=\left(\sigma^{2}N%
\right)^{-1}\tau^{\prime}\tau$).

For the model (\ref{eq:General-Model}), the statistics $Z_{1}$ and $%
Z_{2}^{\prime}Z_{2}$ are independently normal and noncentral Wishart
distributed. Their distributions depend on $\left(\rho,\sigma^{2},\delta_{%
\tau,x}\right)$ and $\left(\rho,\sigma^{2},\omega_{\tau,x}^{2}\right)$,
respectively.

If we condition on the initial observation $y_{1}$, the model (\ref%
{eq:Initial-Model}) for $Y_{T}$ is a special case of (\ref{eq:General-Model}%
), where $x=y_{1}$, $\tau=\eta$, and the regression coefficients are
\begin{equation*}
a\left(\gamma\right)=\rho e_{1}^{\prime}B^{\prime}\text{, }%
b\left(\gamma\right)=1_{T}^{\prime}B^{\prime},\text{ and }%
c\left(\gamma\right)=\sigma.B^{\prime}.
\end{equation*}
Let $\theta_{1}=\left(\rho,\sigma^{2},\delta_{\eta,y_{1}},\omega_{%
\eta,y_{1}}^{2}\right)\in\mathbb{R}^{4}$ be the parameters of the joint
distribution of \eqref{eq:Maximal-Invariant-1}. The likelihood estimator
that maximizes the joint likelihood of \eqref{eq:Maximal-Invariant-1} does
not have the incidental parameter problem since it is parametrized by $%
\theta_{1}$, which has fixed dimension.

If we instead difference out the data, as \citet{Lancaster02} does, the
model (\ref{eq:Model-1-Matrix-Form}) for $\widetilde{Y}_{T}$ is a special
case of (\ref{eq:General-Model}), in which $x$ is absent, the incidental
parameter $\tau $ is $\widetilde{\eta }=\eta -\left( 1-\rho \right) .y_{1}$,
and the regression coefficients are
\begin{equation*}
b\left( \gamma \right) =1_{T}^{\prime }B^{\prime }\text{ and }c\left( \gamma
\right) =\sigma .B^{\prime }.
\end{equation*}

We now show that \citet{Lancaster02}'s differencing method entails
unnecessary efficiency loss. We begin by defining the estimator based on the
likelihood of $\left( Z_{1},Z_{2}^{\prime }Z_{2}\right) $. Let
\begin{equation}
\hat{\theta}_{M1}=\underset{\theta _{1}\in \varTheta}{\arg \min }%
Q_{N}^{M1}\left( \theta _{1}\right) ,  \label{eq:MILEG}
\end{equation}%
where
\begin{eqnarray}
Q_{N}^{M1}\left( \theta _{1}\right) &=&-\frac{1}{2}\ln \sigma ^{2}-\frac{%
\omega _{\eta ,y_{1}}^{2}}{2}-\frac{1}{2\sigma ^{2}}tr\left( \frac{%
DZ_{2}^{\prime }Z_{2}D^{\prime }}{NT}\right) +\frac{1}{2T}\left(
1+A_{1,N}^{2}\right) ^{1/2}  \notag \\
&&-\frac{1}{2\sigma ^{2}NT}\left( Z_{1}D^{\prime }-\left\Vert
y_{1}\right\Vert \left( \rho e_{1}^{\prime }+\delta _{\eta
,y_{1}}1_{T}^{\prime }\right) \right) \left( DZ_{1}^{\prime }-\left\Vert
y_{1}\right\Vert \left( \rho e_{1}+\delta _{\eta ,y_{1}}1_{T}\right) \right)
\notag \\
&&-\frac{1}{2T}\ln \left( 1+\left( 1+A_{1,N}^{2}\right) ^{1/2}\right) ,
\label{eq:Objective-Function-MILE-2}
\end{eqnarray}%
is, up to $o_{p}\left( N^{-1}\right) $ terms, proportional to the
(conditional on $y_{1}$) log-likelihood of $\left( Z_{1},Z_{2}^{\prime
}Z_{2}\right) $ and $A_{1,N}=2\sqrt{\omega _{\eta ,y_{1}}^{2}\frac{%
1_{T}^{\prime }DZ_{2}^{\prime }Z_{2}D^{\prime }1_{T}}{\sigma ^{2}N}}$. The
asymptotic behavior of the estimator $\hat{\theta}_{M1}$ is given next.

\bigskip{}

{\large {}{}{}{}{}{}Lemma }6: Assume that the true parameters $%
\delta_{\eta,y_{1}}$ and $\omega_{\eta,y_{1}}^{2}$ are fixed, respectively,
at $\delta_{\eta,y_{1}}^{\ast}$ and $\omega_{\eta,y_{1}}^{\ast2}>0$ and that
$0<\underset{N\rightarrow\infty}{\lim}\frac{\left\Vert y_{1}\right\Vert ^{2}%
}{N}\equiv\overline{\left\Vert y_{1}\right\Vert }^{2}<\infty$ . As $%
N\rightarrow\infty$ with $T$ fixed,

(A) $\hat{\theta}_{M1}\rightarrow_{p}\theta_{1}^{\ast}=\left(\rho^{\ast},%
\sigma^{\ast2},\delta_{\eta,y_{1}}^{\ast},\omega_{\eta,y_{1}}^{\ast2}%
\right). $

(B) Let $Q_{N}^{M1}\left(\rho,\sigma^{2}\right)$ be the objective function (%
\ref{eq:Objective-Function-MILE-2}) concentrated out of $\delta_{\eta,y_{1}}$
and $\omega_{\eta,y_{1}}^{2}$ and denote the score statistic and the Hessian
matrix by
\begin{equation*}
S_{N}^{M1}\left(\rho,\sigma^{2}\right)=\frac{\partial
Q_{N}^{M1}\left(\rho,\sigma^{2}\right)}{\partial\left(\rho,\sigma^{2}%
\right)^{\prime}}\;\mathrm{and}\;H_{N}^{M1}\left(\rho,\sigma^{2}\right)=%
\frac{\partial^{2}Q_{N}^{M1}\left(\rho,\sigma^{2}\right)}{%
\partial\left(\rho,\sigma^{2}\right)^{\prime}\partial\left(\rho,\sigma^{2}%
\right)},
\end{equation*}
respectively. Also, define the matrix
\begin{equation*}
\mathcal{I}_{T}^{M1}\left(\theta_{1}^{\ast}\right)=\left[%
\begin{array}{cc}
d_{T}^{M1} & -\frac{1_{T}^{\prime}F_{1}^{\ast}1_{T}}{\sigma^{\ast2}T^{2}} \\
-\frac{1_{T}^{\prime}F_{1}^{\ast}1_{T}}{\sigma^{\ast2}T^{2}} & \frac{1}{%
2\left(\sigma^{\ast2}\right)^{2}}\frac{T-1}{T}%
\end{array}%
\right],
\end{equation*}
where
\begin{align*}
d_{T}^{M1} & =\frac{1}{T\left(1+\omega_{\eta,y_{1}}^{\ast2}T\right)}\left(%
\frac{1_{T}^{\prime}F_{1}^{\ast^{\prime}}F_{1}^{\ast}1_{T}}{T}%
+\omega_{\eta,y_{1}}^{\ast2}T\left(\frac{1_{T}^{\prime}F_{1}^{\ast}1_{T}}{T}%
\right)^{2}\right)-\frac{2}{T}\left(\frac{1_{T}^{\prime}F_{1}^{\ast}1_{T}}{T}%
\right)^{2} \\
& +\left(\omega_{\eta,y_{1}}^{\ast2}+\frac{\overline{\left\Vert
y_{1}\right\Vert }^{2}}{\sigma^{\ast2}}\left(\delta_{\eta,y_{1}}^{*}+\rho^{%
\ast}-1\right)^{2}\right)\left(\frac{1_{T}^{\prime}F_{1}^{\ast^{%
\prime}}F_{1}^{\ast}1_{T}}{T}-\left(\frac{1_{T}^{\prime}F_{1}^{\ast}1_{T}}{T}%
\right)^{2}\right) \\
& +\frac{tr\left(F_{1}^{\ast}F_{1}^{\ast^{\prime}}\right)}{T}-\frac{%
1_{T}^{\prime}F_{1}^{\ast^{\prime}}F_{1}^{\ast}1_{T}}{T^{2}}.
\end{align*}
Then, (i) $\sqrt{NT}S_{N}^{M1}\left(\rho^{\ast},\sigma^{\ast2}\right)%
\rightarrow_{d}N\left(0,\mathcal{I}_{T}^{M1}\left(\theta_{1}^{\ast}\right)%
\right)$; (ii) $H_{N}^{M1}\left(\rho^{\ast},\sigma^{\ast2}\right)%
\rightarrow_{p}-\mathcal{I}_{T}^{M1}\left(\theta_{1}^{\ast}\right)$; and
(iii) $\sqrt{NT}\left(%
\begin{array}{c}
\hat{\rho}_{M1}-\rho^{\ast} \\
\hat{\sigma}_{M1}^{2}-\sigma^{\ast2}%
\end{array}%
\right)\rightarrow_{d}N\left(0,\mathcal{I}_{T}^{M1}\left(\theta_{1}^{\ast}%
\right)^{-1}\right)$.

\bigskip{}

Lemma 7 below shows that, in general, \eqref{eq:MILEG} estimates the
structural parameters with strictly smaller variances than the original
estimator that uses differenced data, as defined in \eqref{eq:MILE}.

\bigskip{}

{\large {}{}{}{}{}{}Lemma }7: If $\delta_{\eta,y_{1}}^{\ast}+\rho^{\ast}%
\neq1 $, then \emph{AVar}$\left(\hat{\rho}_{M}\right)>\emph{AVar}\left(\hat{%
\rho}_{M1}\right)$ and \emph{AVar}$\left(\hat{\sigma}_{M}^{2}\right)>\emph{%
AVar}\left(\hat{\sigma}_{M1}^{2}\right)$. If $\delta_{\eta,y_{1}}^{\ast}+%
\rho^{\ast}=1$, then \emph{AVar}$\left(\hat{\rho}_{M}\right)=\emph{AVar}%
\left(\hat{\rho}_{M1}\right)$ and \emph{AVar}$\left(\hat{\sigma}%
_{M}^{2}\right)=\emph{AVar}\left(\hat{\sigma}_{M1}^{2}\right)$.

\section{Random Effects and Moment Conditions\label{Random Effects Sec}}

In Section \ref{Initial Condition Sec} we draw inference on the structural
parameters by conditioning or by differencing the data based on the first
\emph{observed} initial condition $y_{i,1}$. Both methods are fixed-effect
approaches, which do not rely on any further assumptions on the \emph{%
unobserved} data such as $y_{i,0}$. We could instead consider random-effect
approaches which rely on assumptions for the unobserved value $y_{i,0}$.
These include Chamberlain's (correlated)\ random effects and %
\citet{BlundellBond98}'s stationarity assumptions for the first unobserved
value $y_{i,0}$.

In this section, we compare the conditional method with the random-effect
methods using the first moment of
\begin{equation}
\left(x^{\prime}x\right)^{-1}x^{\prime}Y\text{ and }\frac{Y^{\prime}Y}{N}.
\label{eq:Maximal-Invariant-2}
\end{equation}
Define the function
\begin{equation*}
\pi\left(\gamma,\delta_{\tau,x}\right)=a\left(\gamma\right)+\delta_{%
\tau,x}.b\left(\gamma\right).
\end{equation*}
The expectation of the maximal invariant is then given by
\begin{eqnarray*}
E\text{ }\left(x^{\prime}x\right)^{-1}x^{\prime}Y & = &
\pi\left(\gamma,\delta_{\tau,x}\right)\text{ and} \\
E\text{ }\frac{Y^{\prime}Y}{N} & = & \sigma^{2}\left[\pi\left(\gamma,%
\delta_{\tau,x}\right)^{\prime}\omega_{x}^{2}\pi\left(\gamma,\delta_{\tau,x}%
\right)+b\left(\gamma\right)^{\prime}\omega_{\tau,x}^{2}b\left(\gamma\right)%
\right]+c\left(\gamma\right)^{\prime}c\left(\gamma\right),
\end{eqnarray*}

For example, \citet{Lancaster02}'s\ differencing approach yields
\begin{equation*}
E\text{ }\frac{\widetilde{Y}_{T}^{\prime }\widetilde{Y}_{T}}{N}=\sigma
^{2}B_{T}\left[ \omega _{\widetilde{\eta }}^{2}1_{T}1_{T}^{\prime }+I_{T}%
\right] B_{T}^{\prime }.
\end{equation*}%
Hence, we have $\left( T+1\right) T/2$ moments and three unknown parameters:
$\rho $, $\sigma ^{2}$, and $\omega _{\widetilde{\eta }}^{2}$. Conditioning
on $y_{1}$ yields the following (conditional)\ moments based on the maximal
invariant:
\begin{equation}
E\left[ \left. \left( y_{1}^{\prime }y_{1}\right) ^{-1}y_{1}^{\prime
}Y_{T}\right\vert y_{1}\right] =\rho e_{1}^{\prime }B_{T}^{\prime }+\delta
_{\eta ,y_{1}}.1_{T}^{\prime }B_{T}^{\prime }  \label{eq:ConditionalMoment1}
\end{equation}%
and
\begin{eqnarray}
&\hspace{0.01in}&E\left[ \left. \frac{Y_{T}^{\prime }Y_{T}}{N}\right\vert
y_{1}\right] =\sigma ^{2}B_{T}\left\{ \omega _{\eta
,y_{1}}^{2}1_{T}1_{T}^{\prime }+I_{T}\right\} B_{T}^{\prime }
\label{eq:ConditionalMoment2} \\
&\hspace{0.01in}&+\text{ }\sigma ^{2}\left[ \rho B_{T}e_{1}+\delta _{\eta
,y_{1}}.B_{T}1_{T}\right] \omega _{y_{1}}^{2}\left[ \rho e_{1}^{\prime
}B_{T}^{\prime }+\delta _{\eta ,y_{1}}.1_{T}^{\prime }B_{T}^{\prime }\right]
.  \notag
\end{eqnarray}%
The unknown parameters are the autoregressive coefficient $\rho $, the error
variance $\sigma ^{2}$, the OLS coefficient $\delta _{\eta ,y_{1}}$, and the
standardized squared residuals $\omega _{\eta ,y_{1}}^{2}$.

Invariance reduces the information to $T+\left( T+1\right) T/2=\left(
T+1\right) \left( T+2\right) /2-1$ moment conditions which depend on only
four parameters. Conditioning on the first observation yields $T$ more
moments than \citet{Lancaster02}'s differencing method and one additional
parameter (four instead of three). This comparison helps to explain the
efficiency gains from conditioning instead of differencing.

We can then explore the conditional moments (\ref{eq:ConditionalMoment1})
and (\ref{eq:ConditionalMoment2}) to find Minimum Distance (MD) estimators.
We further note that \citet{ArellanoBond91} and \citet{AhnSchmidt95}
implicitly use linear combinations of the moments (\ref%
{eq:ConditionalMoment1}) and (\ref{eq:ConditionalMoment2}) to find moments
for the autoregressive parameter $\rho$.

We instead explore connections between these moments, conditional on the
first observed value $y_{i,1}$, and random effects assumptions for the
unobserved value $y_{i,0}$. To simplify comparison, we continue working with
the model without covariates and with homoskedastic errors.

\subsection{Unobserved Initial Conditions as Incidental Parameters}

As a preliminary to the correlated random effects assumption, consider the
model written recursively to the time $t=0$:
\begin{eqnarray}
Y_{T+1} & = & y_{0}.\rho
e_{1}^{\prime}B_{T+1}^{\prime}+\eta.1_{T+1}^{\prime}B_{T+1}^{\prime}+%
\sigma.U_{T+1}.B_{T+1}^{\prime},\text{ where}  \label{eq:Model-Levels-Matrix}
\\
U_{T+1} & \sim & N\left(0_{N\times\left(T+1\right)},I_{N}\otimes
I_{\left(T+1\right)}\right),  \notag
\end{eqnarray}
and $y_{0}$ is unobserved. This model is again a special case of the linear
setup of \citet{ChamberlainMoreira09} without covariates $x$. The incidental
parameter $\tau=\left[%
\begin{array}{cc}
y_{0} & \eta%
\end{array}%
\right]$ would encapsulate the individual fixed effects and the initial
conditions themselves. The regression coefficients would be given by
\begin{equation*}
b\left(\gamma\right)=\left[%
\begin{array}{c}
\rho e_{1}^{\prime} \\
1_{T+1}^{\prime}%
\end{array}%
\right]B^{\prime}\text{ and }c\left(\gamma\right)=\sigma.B^{\prime}
\end{equation*}
(where we again omit the subscript, here $\left(T+1\right)$, from the
matrices when there is no confusion).

The maximal invariant is $Y_{T+1}^{\prime }Y_{T+1}$, which contains $\left(
T+1\right) \left( T+2\right) /2$ nonredundant moments. Its expectation is
given by
\begin{equation*}
E\text{ }\frac{Y_{T+1}^{\prime }Y_{T+1}}{N}=\sigma ^{2}B_{T+1}\left\{ \left[
\begin{array}{cc}
\rho .e_{1} & 1_{T+1}%
\end{array}%
\right] \omega _{\tau }^{2}\left[
\begin{array}{c}
\rho .e_{1}^{\prime } \\
1_{T+1}^{\prime }%
\end{array}%
\right] +I_{T+1}\right\} B_{T+1}^{\prime }.
\end{equation*}%
The total number of parameters are five: the autoregressive parameter $\rho $%
, the error variance $\sigma ^{2}$, and the three nonredundant elements of $%
\omega _{\tau }^{2}$. The parameters need to be well-behaved as the sample
size $N$ grows, and inference is based on the asymptotic normality of their
respective estimators. If different series start at different points in time
(as typically happens with firms' data), then the parameter $\omega
_{y_{0}}^{2}$ may not be well-behaved or its estimator may not be
asymptotically normal. For example, moment equations depend on terms such as%
\begin{equation*}
\sum_{i=1}^{N}y_{i,0}u_{i,t}.
\end{equation*}%
This term may not be approximately normal if the bulk of observations $%
y_{i,0}$ are close to zero. This happens because the variance ratio of some
terms $y_{i,0}u_{i,t}$ to their sum,%
\begin{equation*}
\max_{j\leq N}\frac{Var\left( y_{j,0}u_{j,t}\right) }{\sum_{i=1}^{N}Var%
\left( y_{i,0}u_{i,t}\right) }
\end{equation*}%
may not be negligible.\footnote{%
The Lindeberg condition holds if and only if the series is approximately
normal and the terms are asymptotically negligible.} This problem is
mitigated by shocks over time, hence conditioning on the observations $%
y_{i,1}$ is more robust.

\subsection{The Correlated Random Effects Estimator}

For us to make a connection to the CRE estimator, we need to include the
vector of ones in $x=1_{N}$. The reason is that the (correlated) random
effects assumption
\begin{equation*}
\left[
\begin{array}{cc}
y_{0} & \eta%
\end{array}%
\right] |x\sim N\left( x.\iota ,I_{N}\otimes \Phi \right) \text{, where }%
\iota =\left( \iota _{1},\iota _{2}\right) .
\end{equation*}%
would need to be invariant to our group of transformations to be decomposed
into an invariant uniform prior and an additional term (and we want to allow
the random effects to have a nonzero mean even without additional
regressors).

The setup is the same as treating the initial condition as an incidental
parameter. However, we include $x=1_{N}$ and define the regression
coefficient on the vector of ones, as in Section 7 of %
\citet{ChamberlainMoreira09}. The maximal invariant is
\begin{equation*}
1_{N}^{\prime}Y_{T+1}\text{ and }Y_{T+1}^{\prime}Y_{T+1}.
\end{equation*}
Their (conditional on $\tau=\left[%
\begin{array}{cc}
y_{0} & \eta%
\end{array}%
\right]$) expectation is given by
\begin{eqnarray*}
E\left[1_{N}^{\prime}Y_{T+1}\right] & = & \delta_{\tau,1_{N}}\left[%
\begin{array}{c}
\rho.e_{1}^{\prime} \\
1_{T+1}^{\prime}%
\end{array}%
\right]B_{T+1}^{\prime}\text{ and} \\
E\left[\frac{Y_{T+1}^{\prime}Y_{T+1}}{N}\right] & = & B_{T+1}\left\{ \left[%
\begin{array}{cc}
\rho.e_{1} & 1_{T+1}%
\end{array}%
\right]\left(\sigma^{2}\omega_{\tau,1_{N}}^{2}+\delta_{\tau,1_{N}}^{\prime}%
\delta_{\tau,1_{N}}\right)\left[%
\begin{array}{c}
\rho.e_{1}^{\prime} \\
1_{T+1}^{\prime}%
\end{array}%
\right]+\sigma^{2}.I_{T+1}\right\} B_{T+1}^{\prime}.
\end{eqnarray*}
The unknown parameters are the autoregressive coefficient $\rho$, the error
variance $\sigma^{2}$, the sample averages $\delta_{\tau,1_{N}}$, and
standardized squared deviations $\omega_{\tau,1_{N}}^{2}$. Hence, the
invariance argument reduces the data space to $T+1+\left(T+1\right)\left(T+2%
\right)/2$ moment conditions, which depend on seven parameters. The
parameter $\omega_{\tau,1_{N}}^{2}$ may not be well-behaved and its
estimator may not be asymptotically normal.

Under the random effects assumption, the model becomes

\begin{equation*}
Y_{T+1}=1_{N}.a\left(\gamma\right)+U_{T+1}.c\left(\gamma\right),
\end{equation*}
where the coefficients are given by
\begin{eqnarray*}
a\left(\gamma\right) & = & \iota\left[%
\begin{array}{c}
\rho e_{1}^{\prime} \\
1_{T+1}^{\prime}%
\end{array}%
\right]B_{T+1}^{\prime}\text{ and} \\
c\left(\gamma\right)^{\prime}c\left(\gamma\right) & = & B_{T+1}\left\{ \left[%
\begin{array}{cc}
\rho.e_{1} & 1_{T+1}%
\end{array}%
\right]\Phi\left[%
\begin{array}{c}
\rho e_{1}^{\prime} \\
1_{T+1}^{\prime}%
\end{array}%
\right]+\sigma^{2}I_{T+1}\right\} B_{T+1}^{\prime}.
\end{eqnarray*}

The (unconditional) expectation of the maximal invariant has the same
functional form as the moments conditional on $\tau=\left[%
\begin{array}{cc}
y_{0} & \eta%
\end{array}%
\right]$:
\begin{eqnarray*}
E\left[\frac{1_{N}^{\prime}Y_{T+1}}{N}\right] & = & \iota\left[%
\begin{array}{c}
\rho.e_{1}^{\prime} \\
1_{T+1}^{\prime}%
\end{array}%
\right]B_{T+1}^{\prime}\text{ and} \\
E\left[\frac{Y_{T+1}^{\prime}Y_{T+1}}{N}\right] & = & B_{T+1}\left\{ \left[%
\begin{array}{cc}
\rho.e_{1} & 1_{T+1}%
\end{array}%
\right]\left(\Phi+\iota^{\prime}\iota\right)\left[%
\begin{array}{c}
\rho.e_{1}^{\prime} \\
1_{T+1}^{\prime}%
\end{array}%
\right]+\sigma^{2}.I_{T+1}\right\} B_{T+1}^{\prime}.
\end{eqnarray*}
So the criticism on lack of robustness to the data-generating process for $%
y_{i,0}$ is applicable here as well.

\subsection{Blundell and Bond (1998)}

\citet{BlundellBond98} instead consider a different assumption, that $y_{0}$
does not deviate systematically from the stationary mean $%
\eta/\left(1-\rho\right)$. We start with the assumption
\begin{equation*}
y_{0}\sim N\left(\frac{\eta}{1-\rho},\sigma_{0}^{2}.I_{N}\right).
\end{equation*}
We assume normality for the purposed of invariance. However, only the mean
and variance are relevant for the expectation calculations derived below.

Under this assumption, the model is equivalent to
\begin{equation*}
Y_{T+1}=\eta.b\left(\gamma\right)+U_{T+1}.c\left(\gamma\right),
\end{equation*}
where the coefficients are given by
\begin{eqnarray*}
b\left(\gamma\right) & = & 1_{T+1}^{\prime}B_{T+1}^{\prime}+\frac{\rho}{%
1-\rho}e_{1}^{\prime}B_{T+1}^{\prime}\text{ and} \\
c\left(\gamma\right)^{\prime}c\left(\gamma\right) & = &
B_{T+1}\left(\sigma_{0}^{2}\rho^{2}e_{1}e_{1}^{\prime}+\sigma^{2}I_{T+1}%
\right)B_{T+1}^{\prime}.
\end{eqnarray*}
The maximal invariant is given by $Y_{T+1}^{\prime}Y_{T+1}$:
\begin{eqnarray*}
& \hspace{0.01in} & E\text{ }\frac{Y_{T+1}^{\prime}Y_{T+1}}{N}%
=B_{T+1}\left(\sigma_{0}^{2}\rho^{2}e_{1}e_{1}^{\prime}+\sigma^{2}I_{T+1}%
\right)B_{T+1}^{\prime} \\
& \hspace{0.01in} & +\sigma^{2}\omega_{\eta}^{2}B_{T+1}\left[%
1_{T+1}1_{T+1}^{\prime}+\frac{\rho}{1-\rho}\left(e_{1}1_{T+1}^{%
\prime}+1_{T+1}e_{1}^{\prime}\right)+\frac{\rho^{2}}{\left(1-\rho\right)^{2}}%
e_{1}e_{1}^{\prime}\right]B_{T+1}^{\prime}.
\end{eqnarray*}
This expectation depends on four parameters: $\rho$, $\sigma^{2},$ $%
\omega_{\eta}^{2}$, and $\sigma_{0}^{2}$.

We can re-arrange some of these $\left(T+1\right)(T+2)/2$ moments so that
the expectation is zero at the true autoregressive coefficient $\rho$. For
example,
\begin{equation}
E\left[\left(y_{i,2}-y_{i,1}\right).\left(y_{i,3}-\rho y_{i,2}\right)\right]%
=0.  \label{eq:BB98}
\end{equation}
As \citet{BlundellBond98} also note, this expectation may fail to be zero if
the stationarity assumption breaks down. For example, this moment condition
breaks down if either $y_{i,1}=k$ or even if $y_{i,1}\overset{iid}{\sim}%
N\left(0,\sigma_{0}^{2}\right)$.

It is worthwhile making a connection between the stationarity assumption and
inference conditional on the first observations. Since inference is
conditional on $y_{1}$, we could stack the quantities together and look at
the (conditional) expectation of
\begin{equation*}
Y_{T+1}^{\prime}Y_{T+1}=\left[%
\begin{array}{cc}
y_{1}^{\prime}y_{1} & y_{1}^{\prime}Y_{T} \\
Y_{T}^{\prime}y_{1} & Y_{T}^{\prime}Y_{T}%
\end{array}%
\right].
\end{equation*}
As in the derivation based on that of \citet{BlundellBond98}, we have
exactly the same quantity $Y_{T+1}^{\prime}Y_{T+1}$. However, we have five
parameters (if we consider $y_{1}^{\prime}y_{1}$ itself to be the
parameter). Alternatively, we lose one moment (from removing $%
y_{1}^{\prime}y_{1}$ itself) and four parameters. If the stationarity
assumption is correct, there should be efficiency loss from making inference
conditional on the first observation. On the other hand, conditional
inference should be robust to different data-generating process for the
initial data values.

\section{Conclusion}

This paper studies the Bayesian estimator of \citet{Lancaster02}, which is
invariant to natural rotations of the data. The likelihood of the maximal
invariant can be divided into two parts, one of which is maximized to obtain
Lancaster's estimator. The second part is not asymptotically negligible and,
as such, can be used to attain a more efficient estimator. A natural
conclusion is that it is unnecessary to use the data to choose the correct
inflection point of Lancaster's score likelihood. \citet{Lancaster02}'s
theory is based on differencing the data using the first observation.
Instead, we can condition on the first observation itself to further improve
efficiency. The conditional argument is essentially a fixed-effects approach
in which we make no further assumptions on unobserved data.

Current practice in economics uses standard GMM methods based on data
differencing; e.g., \citet{AthanasoglouBrissimisDelis08} on bank
profitability, \citet{GuisoPistaferriSchivardi05} on risk allocation, %
\citet{KoningsVandenbussche05} on antidumping protection effects, and %
\citet{TopalovaKhandelwal11} on tariffs' change on firm productivity, among
others. Instead, we advocate using moments based on invariant statistics
from a model that allows for heteroskedasticity and factor structure. It is
relatively straightforward to extend our method to allow for covariates in
the linear model of \citet{ChamberlainMoreira09}. \citet{Bai13} allows for
general time-series heteroskedasticity as opposed to assuming specific
moving average (MA) processes and data values lagged enough as instruments. %
\citet{ChamberlainMoreira09} and \citet{MoonWeidner15} suggest including
factors which generalize individual and time effects. Conditioning on the
first observation can then provide reliable inference in dynamic panel data
models using moments based on invariant statistics for more general models.

\addcontentsline{toc}{section}{\refname}
\bibliographystyle{econometrica}
\bibliography{References}

\end{document}